%% file: paper1526.tex
\documentclass[runningheads]{llncs}
\usepackage{graphicx}
\usepackage{amsmath}
\usepackage{amsfonts}
\usepackage{bigints}
\usepackage{IEEEtrantools}
\usepackage{booktabs}
\usepackage{tabularx}
\usepackage{caption3}
\usepackage{color}
\usepackage{subfigure}
\newcolumntype{Y}{>{\centering\arraybackslash}X}

\input{math_commands}

\begin{document}
\title{ Self-Supervised Vessel Enhancement Using Flow-Based Consistencies}
%
%
%

\author{Rohit Jena \inst{1} \and
Sumedha Singla \inst{2} \and
Kayhan Batmanghelich \inst{2}
}

\institute{Carnegie Mellon University, Pittsburgh, PA, USA \and
University of Pittsburgh, PA, USA 
}
\authorrunning{Jena et al.}


%
%
\maketitle              
\begin{abstract}
\input{sections/abstract_v3}

\end{abstract}

\input{sections/intro_v2}

\input{sections/method_v3}
\input{sections/experiments_v2}

\input{sections/conclusion_v1_s}

\bibliographystyle{splncs04}
\bibliography{references}
\clearpage
\input{sections/appendix_v1}

\end{document}

%% file: math_commands.tex
\usepackage{amsmath,amsfonts,bm}

\def\eqref#1{eq.~\ref{#1}}

\def\1{\bm{1}}

\def\vb{{\bm{b}}}

\def\vp{{\bm{p}}}
\def\vq{{\bm{q}}}

\def\vu{{\bm{u}}}

\DeclareMathAlphabet{\mathsfit}{\encodingdefault}{\sfdefault}{m}{sl}
\SetMathAlphabet{\mathsfit}{bold}{\encodingdefault}{\sfdefault}{bx}{n}

\def\sR{{\mathbb{R}}}
\def\sS{{\mathbb{S}}}



%% file: sections/abstract_v3.tex
Vessel segmentation is an essential task in many clinical applications. 
Although supervised methods have achieved state-of-art performance, acquiring expert annotation is laborious and mostly limited for two-dimensional datasets with a small sample size. 
On the contrary, unsupervised methods rely on handcrafted features to detect tube-like structures such as vessels. 
However, those methods require complex pipelines involving several hyper-parameters and design choices rendering the procedure sensitive, dataset-specific, and not generalizable.
We propose a self-supervised method with a limited number of hyper-parameters that is generalizable across modalities.  
Our method uses tube-like structure properties, such as connectivity, profile consistency, and bifurcation, to introduce inductive bias into a learning algorithm.
To model those properties, we generate a vector field that we refer to as a \emph{flow}. 
Our experiments on various public datasets in 2D and 3D show that our method performs better than unsupervised methods while learning useful transferable features from unlabeled data.
Unlike generic self-supervised methods, the learned features learn vessel-relevant features that are transferable for supervised approaches, which is essential when the number of annotated data is limited.

%% file: sections/intro_v2.tex
\section{Introduction}
Tube-like structures, such as vessels and airways, are ubiquitous in studying human anatomy. 
Segmenting such structures is essential for characterizing the progression of many diseases~\cite{estepar2013computed,junior2013automatic}.
Supervised deep learning methods have made significant progress for accurate segmentation~\cite{guo2020sa,tetteh2018deepvesselnet,SoomroReview2019}, but annotated datasets are largely limited to 2D data and often have a small sample size.
We develop a self-supervised task that incorporates the structure's key properties and learns optimal representation for the structure.
Our method is applicable for both 2D and 3D, and it can be employed to bootstrap supervised methods when the number of \textit{annotated} data is limited.

Automatic vessel segmentation is a challenging problem, given that the vascular networks are complex multi-level tree structures with high variability in local geometry, curvature, and radius, which further varies across modalities and subjects.
Recently, various deep learning (DL) based techniques have been proposed for various segmentation tasks of tube-like structures~\cite{guo2020sa,tetteh2018deepvesselnet,SoomroReview2019}.
Training supervised DL algorithms requires many annotated images, which is particularly laborious for complex tube-like structures.
Due to the lack of a large-scale annotated dataset to train a supervised method,  \emph{unsupervised}  vessel segmentation methods are still popular, and there is a growing interest in deploying {DL-based} \emph{unsupervised} and \emph{self-supervised} methods~\cite{taleb20203d,zhou2019models}.

An unsupervised pipeline specially designed for vessel segmentation varies across modalities and image dimensionality (e.g., 2D retinography~\cite{fraz2012blood} and 3D thoracic CT~\cite{rudyanto2014comparing}). 
Most state-of-the-art pipelines rely on hand-crafted features based on different variants of classical Hessian-based scale-space filters~\cite{frangi,hessian,sato}. Achieving state-of-the-art results requires post-processing using different techniques such as particle sampling~\cite{estepar2013computed} and region growing~\cite{yu2008multiscale} with several design choices for each step with their corresponding hyper-parameters. 
Such design renders the procedure problem-specific and not transferable across domains. In contrast, we propose an end-to-end unsupervised vessel segmentation model that generalizes across modalities and dimensions. 

Our work is inspired by the matched filter response (MFR)  method~\cite{chaudhuri1989detection} and scale-space approaches \cite{estepar2012computational}. 
These methods  model the vessel as piece-wise linear segments and use multiple Gaussian kernels to identify vessel-like structures. 
Our approach is a modern adaptation of MFR and scale-space using a fully convolutional network (FCN).
This paper's central idea is to model the vessel with a \emph{flow} which defines a continuous path;  the profile of the tube-like structure matches with expected \emph{template} as we \emph{walk} along with the flow.
We use the notion of walking along with the flow as an inductive bias for a self-supervised method that learns the set of optimal features.
The FCN architecture, which is used to infer the flow, naturally processes the image in a multi-scale fashion. 

The paper makes the following contributions. (1) It proposes a  self-supervision inspired by the problem, in this case, segmenting tube-like structures. (2) The method is generic and can be deployed for 2D and 3D images for enhancing vessel in various modalities. (3) Unlike other unsupervised methods using fixed hand-crafted features, our method adapts the features as the dataset changes. 
When annotated data is limited, the trained features can also be transferred to boost a supervised method's performance.
To the best of our knowledge, our work is the first unsupervised deep learning method that takes a raw image as input and outputs per-pixel vessel statistics as output, along with an associated per-pixel vesselness score.
We evaluate the performance of our method on real 2D  and 3D datasets. 
We also show the efficacy of incorporating context into self-supervised learning by comparing our method with state-of-the-art self-supervised pretraining tasks.

%% file: sections/method_v3.tex

\section{Method}

We use some of the key properties of \emph{tube-like} structures (\emph{e.g.,} vessels) to define a self-supervised algorithm. We use vessel structure as a running example, however, the proposed method is general and can be applied to other tube-like structures such as airways. We consider the following three properties:

%
    %
\textbf{(1) Path Continuity}: The path continuity assumes that the vessel trees' structure and arrangement can be viewed as a continuous path in space (represented by a \textit{vector field}).
Each vector in the \emph{vector field} indicates the radius and direction of the vessel.
We refer to this vector field as \emph{vessel flow} field.
\textbf{(2) Profile Consistency}: A Tube-like structure has similar profiles at different points when the planes are perpendicular to the structure's medial axis (i.e., the vector field). 
In other words, the orthogonal profile of the tube-like structure can be approximated by a predefined template, $T$.
For a vessel, $T$ is simply a unit disk.
\textbf{(3) Bifurcation}: The possible bifurcation refers to the fact that the vessel may split into two sub-vessels. We use two vector fields to model the bifurcation. Wherever there is a bifurcation, the two vector fields point toward each bifurcation branch; otherwise, they are aligned with the vessel flow.

\begin{figure}[!t]
    \centering
    \includegraphics[width=0.88\textwidth]{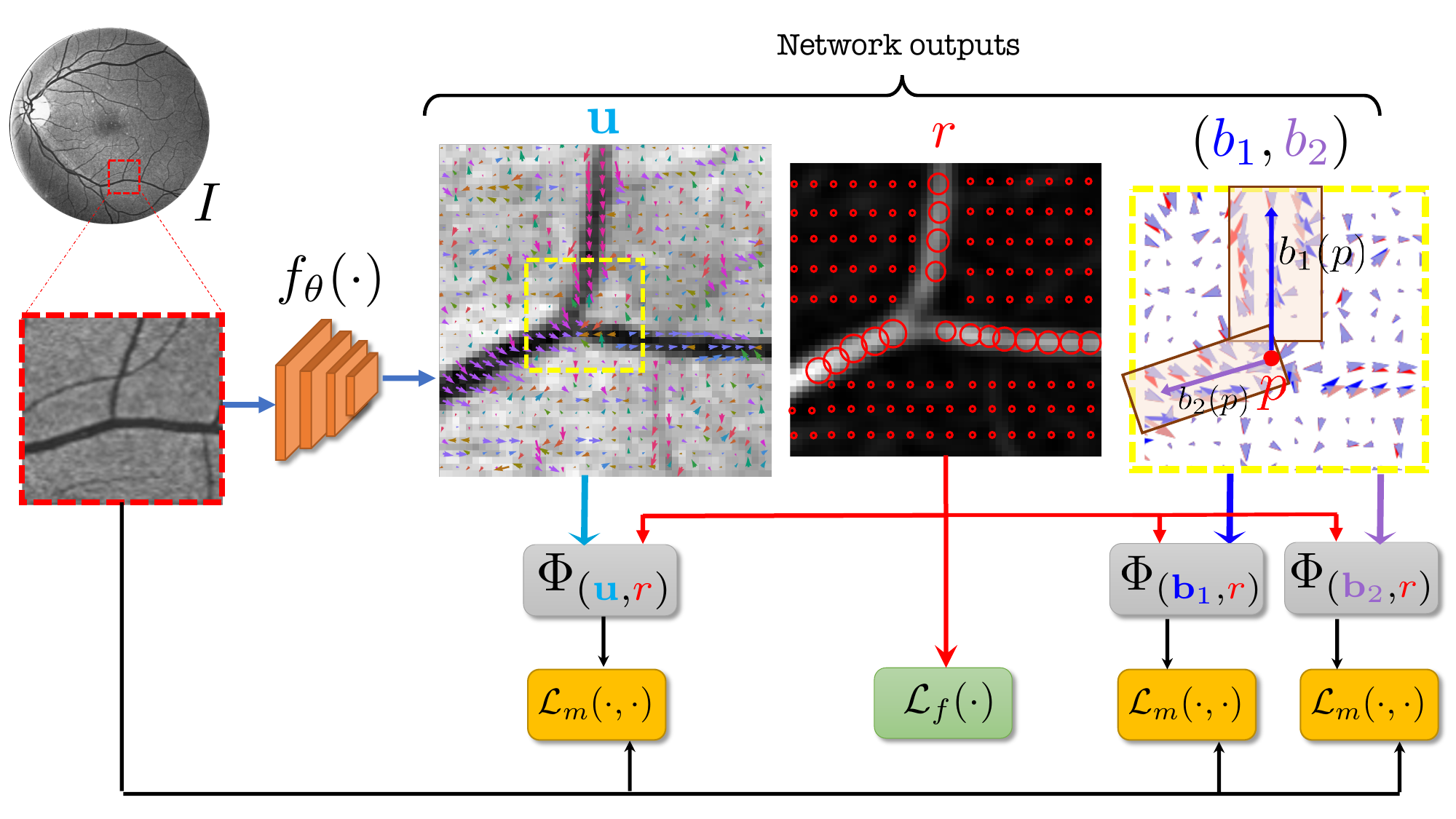}
    \caption{The overall framework of the proposed method. The network $f_{\theta}$ generates a scalar field $r$ specifying the local width of the vessel, vector field $\vu$ denoting the direction of the vessel, and a tuple of two vector fields $\vb_1,\vb_2$ pointing towards branches in cases of bifurcation.
    $\Phi_{(\vu, r)}$ is a transformation parameterized by a flow field $\vu$ and a radius field $r$ that is composed with the image $I$.
    There are two losses, $\mathcal{L}_m(\cdot,\cdot)$, and $\mathcal{L}_f(\cdot)$  encouraging the properties of the tube-like structures.
    } 
    \label{fig:framework}
\end{figure}

\textbf{General Framework:}
Let $I: \Omega \rightarrow \mathbb{R}$ represent input image defined over the $d$-dimensional domain $\Omega \subseteq \sR^d$. We consider $d \in \{2, 3\}$. We assume a network $f_\theta$ generates three outputs: $\vu$, $r$, and $(\vb_1, \vb_2)$.
Let $\vu : \Omega \to \sS^{d-1}$ denote  the vessel flow, defined as a vector field where for any point $\vp \in \Omega$, where $\sS^{d-1}$ is a $d$-dimensional unit hypersphere. 
We define $\vb_1: \Omega \rightarrow \sS^{d-1}$ and $\vb_2: \Omega \rightarrow \sS^{d-1}$ to represent the two directions of the bifurcation.
Furthermore, $r: \Omega \to \sR^{+} $ is defined as a scalar field to represent the radius of the vessel.
The idea has been shown in \figurename{\ref{fig:framework}}. We define the losses to enforce those properties. We first explain the profile symmetry followed by path continuity and bifurcation.

\subsection{Profile Consistency}
The vessel flow, $\vu$, specifies the directionality of the vessel.
At each point $\vp$, the orthogonal plane to the vessel flow specifies the vessel's profile.
Given a template $T$, one expects the resized vessel profile to match the template.
We use $I \circ \Phi_{(\vu , r)}(\vp)$ to denote the profile of image $I$ at point $\vp$ along the direction $\vu(\vp)$.
The output $r(\vp)$ 
specifies the radial field of view of the profile.
The image is resized by resampling according to $r(\vp)$.
The $\Phi_{(\vu , r)}$ denotes this transformation and $\circ$ denotes the composition of the transformation with image $I$.  
Similar to image registration, we maximize a similarity metric ($\mathcal{S}(\cdot,\cdot)$) between the template ($T$) and the transformed profile ($I \circ \Phi_{(\vu , r)}(\vp)$) over the entire domain ($\Omega$):
\begin{IEEEeqnarray}{c}
     \mathcal{L}_m(\vu,r; I,T) = - \int_{\Omega}  \overbrace{ \mathcal{S}( I \circ \Phi_{(\vu , r)}(\vp) , T)}^{V(\vp)} d\vp.
     \label{eq:alignment}
\end{IEEEeqnarray}
Various choices are possible for $\mathcal{S}(\cdot,\cdot)$. We use Normalized Cross-Correlation (NCC) because of its robustness to changes in the illumination and contrast. The $\mathcal{L}_m(\vu,r; I,T)$ indicates that the value of the loss is a function of vessel flow, $\vu$ and vessel radius, $r$ given the image and template. For other tube-like structures such as airways, one can change the template.
We define $V(\vp)$ in the domain as a \textbf{vesselness} value for the pixel $\vp$.

\subsection{Path Continuity}
We impose vessel flow continuity for two reasons.
First, we assume the entire vessel structure is a connected component.
Second, for a point $\vp$, the vessel flow $\vu(\vp)$ and $-\vu(\vp)$ result in the same vessel profile.
Such sign ambiguity may result in discontinuities in vector field $\vu$.
To prevent this, we propose an alignment loss that encourages $\vu$ to have a consistent direction along the vessel.
We do that by \emph{walking} along the direction of the vessel flow.
We formulate the walk by a stationary path integral.
For a point $\vp$, let's assume $\vq_\vp(t)$ is the new coordinate after walking for $t$ from $\vp$.
The flow vector at the final point is $\vu(\vq_\vp(t))$. If the vector field is consistent and continous, the inner product between $\vu(\vp)$ and $\vu(\vq_\vp(t))$ should be large; i.e., they should be almost aligned. We define a loss to maximize that loss over the entire domain:
\begin{IEEEeqnarray}{cc}
\dot{\vq}_\vp = \vu(\vp), & \vq_\vp(t=0) = \vp, \nonumber \\
\mathcal{L}_f(\vu, r) &= - \int_{\Omega} \int_{0}^{2r(\vp)} \left< \vu(\vp), \vu(\vq(t)) \right> dt d\vp,
\end{IEEEeqnarray}

where $\vq$ is the position of the walk, $\dot{\vq}$ is the time derivative and the given point, $\vp$ is the initial point in the path, and $\left< \cdot, \cdot \right>$ represents inner product.
We set $2r(\vp)$ as an upper limit in the integral so that the length of the traversal path is proportional to the size of the vessel.

\subsection{Bifurcation}
A bifurcation point (BP) is where the main vessel splits into two branches.
Similar to vessel flow $\vu$, we predict bifurcation flow fields $\vb_1$ and $\vb_2$ as vector fields representing the two directions of the bifurcation. Although the BP does not match the canonical template $T$, the incoming vessel and the two branches should match $T$. We define the birfurcation loss as an extension of $\mathcal{L}_m$,
\begin{equation}
\mathcal{L}_b(\vb_1, \vb_2,r; I,T) = \mathcal{L}_m(\vb_1,r; I,T)  + \mathcal{L}_m(\vb_2,r; I,T). 
\label{eq:bifurc}
\end{equation}
Note that, in the absence of bifurcation, $\vb_1(\vp) = \vb_2(\vp) = -\vu(\vp)$ minimizes the same loss as $\mathcal{L}_m$ in the opposite direction of the vessel flow. 
Hence in practice, we can add the loss function in Eq. \ref{eq:bifurc} to Eq. \ref{eq:alignment}. 

\begin{table*}[!t]
    \centering
    \renewcommand{\arraystretch}{1.27}
    \resizebox{0.88\textwidth}{!}{%
    \begin{tabularx}{\textwidth}{@{}p{2cm}YYYY|YYYY@{}} \toprule
         &
         \multicolumn{4}{c}{\textbf{DRIVE}} & 
         \multicolumn{4}{c}{ \textbf{STARE} } \\ \hline
         \textbf{Method} 
         & \textbf{AUC} & \textbf{Acc} &  \textbf{LAcc}  & \textbf{Dice} & \textbf{AUC} & \textbf{Acc} & \textbf{LAcc} & \textbf{Dice}\\ \hline
         Hessian & 0.55 & 52.90 & 60.31 &  0.25 & 0.53 & 46.84 & 59.24 &  0.21  \\
         Frangi & 0.93 &	94.27 &	72.02  &	 0.67	& 0.93  &	91.16 &	67.12  &	 0.59 \\ 
         Sato & 0.94 & 93.67 & 71.90   & 0.66 & 0.94 & 91.03 & 66.52 &   0.58 \\
         Meijering  & 0.94 & 93.83 & 72.70  &  0.67 & 0.94 & 90.29 & 68.11   & 0.58 \\
         Ours  & \textbf{0.96} & \textbf{95.69} & \textbf{75.83 }  &  \textbf{0.74} &\textbf{0.96} &	\textbf{95.29} &	\textbf{69.44} & \textbf{0.68}\\   \hline
         &\multicolumn{4}{c}{\textbf{HRF}} & 
         \multicolumn{4}{c}{ \textbf{RITE} } \\ \hline
         Hessian & 0.37 & 33.70 & 55.35 & 0.17 & 0.51 & 49.77 & 60.21 & 0.23 \\ 
         Frangi & 0.93 & 93.90 & 70.59 & 0.63 & 0.94 & 94.26 & 73.10 & 0.68 \\ 
         Sato & 0.94 & 93.73 & 70.25 & 0.62 & 0.95 & 94.01 & 72.86 & 0.67 \\ 
         Meijering  & 0.92 & 91.11 & 72.73 & 0.58 & 0.95  & 94.77 & \textbf{77.32} & 0.71 \\ 
         Ours  & \textbf{0.95} & \textbf{94.96} & \textbf{72.78} & \textbf{0.68} & \textbf{0.97} & \textbf{96.07} & 77.05 & \textbf{0.75} \\
         \bottomrule
    \end{tabularx}
}
    \caption{Evaluation of unsupervised vessel segmentation on 2D datasets. Acc is the global accuracy over the entire image, LAcc is the local accuracy around the dilated vessel regions defined in \cite{localacc}. There are the significant improvements in Dice score, which is a crucial metric in sparse segmentation.}
    \label{tab:unsupervised2d}
\end{table*}


\subsection{Implementation Details}
The overall cost function is as follows:
\begin{IEEEeqnarray}{lll}
  \min_{\theta} & \quad & \mathcal{L}_m(\vu,r; I,T) + \lambda_1 \mathcal{L}_f(\vu, r) + \lambda_2  \mathcal{L}_b(\vb_1, \vb_2,r; I,T) \nonumber \\ 
  \text{s.t:} & \quad &(r,\vu, \vb_1, \vb_2) = f_{\theta}(I),
\end{IEEEeqnarray}
where $\lambda_i$'s are weighting hyper-parameters and $f_{\theta}(\cdot)$ is the network. To be consistent with the literature, we adopt the U-Net architecture for $f_{\theta}$~\cite{retinaunet}.
The models are trained with the Adam optimizer, with a learning rate of $0.001$ for 2D and $0.0003$ for 3D.
For 2D images, we use the entire image as input, with a batch size of 4, and for 3D, we use a $64\times 64 \times 64$ patch with a batch size of 1.
During training, we augment the images by flipping and rotating in increments of 90 degrees.
All our models are trained on NVIDIA Tesla V100-SXM2 GPUs.

%% file: sections/experiments_v2.tex
\section{Experiments}

\begin{figure}[!t]
    \centering
    \includegraphics[width=0.31\textwidth]{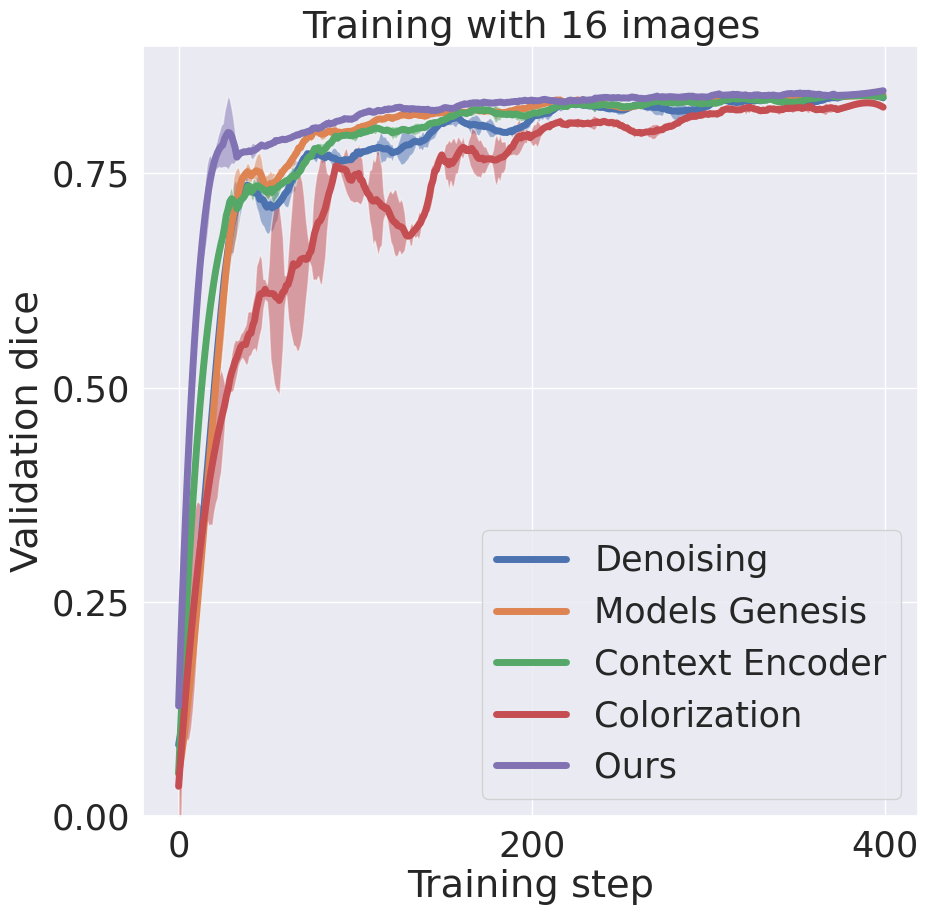}
    \includegraphics[width=0.31\textwidth]{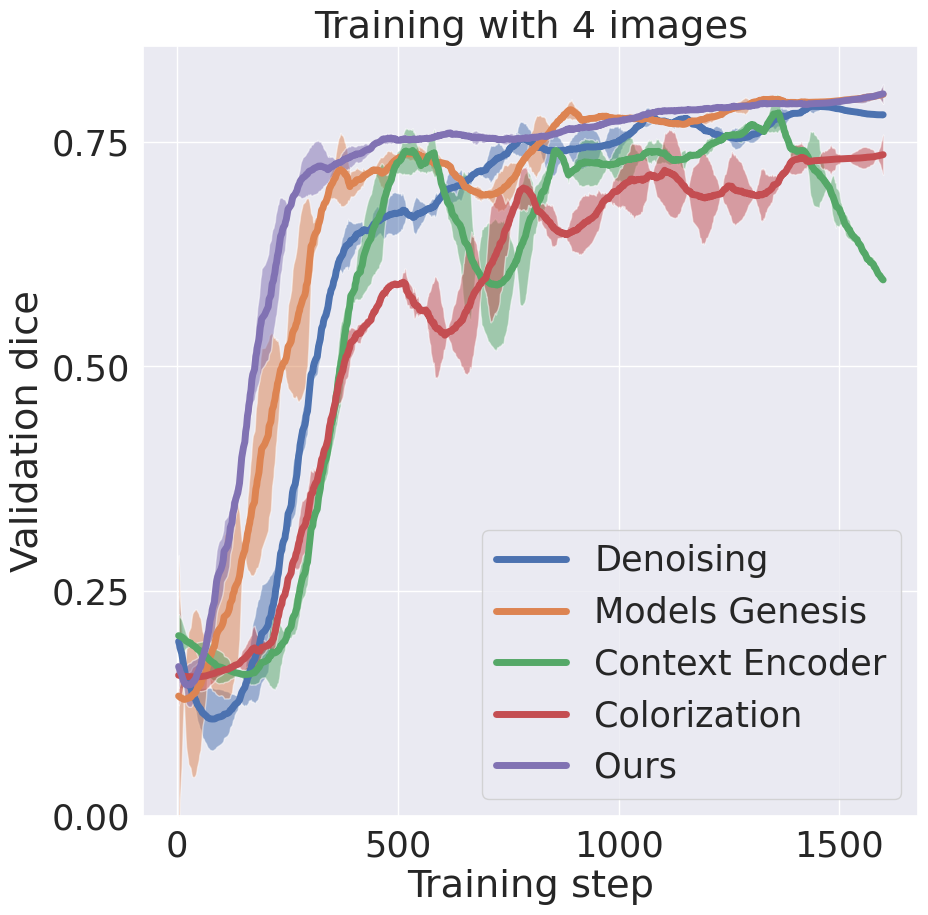}
    \includegraphics[width=0.34\textwidth]{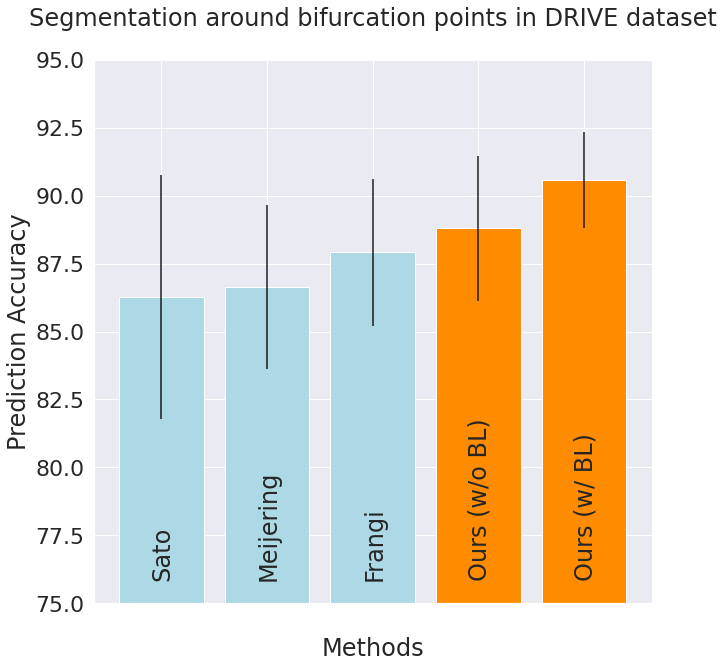}
    \resizebox{0.95\textwidth}{!}{
    \begin{minipage}{0.32\textwidth}
    \centering
    (a) 
    \end{minipage}
    \begin{minipage}{0.32\textwidth}
    \centering
    (b) 
    \end{minipage}
    \begin{minipage}{0.32\textwidth}
    \centering
    (c) 
    \end{minipage}
    }
    \caption{Training curves for supervised vessel-segmentation on STARE dataset with \textbf{(a) }limited data (4 images), \textbf{(b)} more data (16 images), after self-supervised pretraining on DRIVE and \textbf{(c)} Bifurcation segmentation performance on DRIVE dataset.
    }
    \label{fig:selfsupervised}
\end{figure}

We perform three experiments to evaluate our method.
(1) We compare the performance of our approach with commonly used unsupervised methods for vessel segmentation: Frangi~\cite{frangi}, Sato~\cite{sato}, Hybrid Hessian~\cite{hessian} and Meijering~\cite{meijering} filters on four 2D datasets and two 3D datasets, all of which are publicly available.  
(2) We study the efficiency of learned representation for the downstream vessel segmentation task.
To do that, we compare our method with existing self-supervised methods.
(3) We examine the efficacy of the bifurcation loss in segmenting the regions around bifurcation points in 2D compared to methods which do not consider bifurcations.

\textbf{Datasets:} For the 2D experiments, we use four publicly available retinal image datasets: DRIVE~\cite{niemeijer2004comparative}, STARE~\cite{stare}, HRF ~\cite{HRF} and RITE ~\cite{RITE}.
The DRIVE and RITE datasets consist of 40 images, divided into a training and testing set of 20 images.
The STARE database consists of 20 images, each image with two sets of segmented images.
The HRF dataset consists of 45 retinopathy images which we divide into a training set of 21 images and testing set of 24 images.
For 3D vessel segmentation, we use the VESSEL12 dataset~\cite{VESSEL12} consisting of 20 CT lung images from a variety of sources. 
The dataset also contains 3 images with sparsely annotated vessel and non-vessel locations along 3 axial slices, which we use as a test set.
We also use the TubeTK dataset~\cite{tubetk} which consists of 3D MRA images of 100 healthy patients of size $448\times448\times128$. We use 42 images with ground truths as the test set, and the remaining images are used for training the network. 

\subsection{Comparison with Unsupervised Methods}
We compare our model against popular vessel enhancement methods, including Frangi, Sato, Hessian, and Meijering filters.  
All the methods take raw images as input and produce a vessel-enhanced image as output. 
The enhanced image is segmented into a binary map using a hard threshold, which is selected to achieve a maximum dice score over the training dataset. 
The performance is then reported over the test set using the hard threshold.
Table~\ref{tab:unsupervised2d}  presents a quantitative comparison of different vessel-segmentation methods as a binary classification problem. 
We reported results on five measures, namely, the area under the curve (AUC) of ROC curves, accuracy (acc), local-accuracy (LAcc)~\cite{localacc}, and dice score.


For the VESSEL12 dataset, we drop the Dice score since we do not have access to a dense ground truth.
Therefore, we treat the problem as a classification problem and compare sensitivity (sens) and specificity (spec) as well.
For TubeTK, we compare the Dice score of the methods with the dense ground truth.  
The results are summarized in Table \ref{tab:datasets3d}.
Similar to 2D, our method performs consistently across datasets, and has a significantly higher Dice score. 
     %
Without using any annotations for training, our method outperforms other commonly used unsupervised vessel enhancement methods.
Since vessel segmentation is a sparse segmentation problem, the critical metrics are Dice score and Local Accuracy, on which our method has a significant improvement over baselines.

\begin{table}[!t]
    \centering
    \renewcommand{\arraystretch}{1.2}
    \resizebox{0.9\textwidth}{!}{%
    \begin{tabularx}{0.99\textwidth}{XYYYY|YYY} \toprule
        & \multicolumn{4}{c}{\textbf{VESSEL12}} 
        & \multicolumn{3}{c}{\textbf{TubeTK}}  \\ \hline
        \textbf{Method} & \textbf{Acc} & \textbf{Spec} & \textbf{Sens} & \textbf{AUC} & \textbf{Acc} & \textbf{AUC} & \textbf{Dice} \\ \hline
        Sato	& 79.1	& 0.81	& 0.74	& 0.88 & 91.17 & 0.74 & 0.15 \\
        Meijering	& 90.16	& 0.89	& 0.92	& 0.96 & 97.25 & 0.83 & 0.34 \\
        Frangi	& \textbf{96.88}	& \textbf{0.97}	& 0.96	& 0.97 & 98.79 & 0.90 & 0.42\\
        Ours	& 95.49	& 0.92	& \textbf{0.99}	& \textbf{0.99} & \textbf{99.05} & \textbf{0.95} & \textbf{0.59} \\ \bottomrule
    \end{tabularx}
    }
    \caption{Results on the VESSEL12 and TubeTK test images.
    Our method has significant improvement in Dice score for TubeTK, which is a critical metric in sparse segmentation.
    Our method also compares well with Frangi on a sparsely annotated ground truth.
    }
    \label{tab:datasets3d}
\end{table}

\subsection{Efficacy of the Representation}
Since our method is learning based, it can learn feature representations that are essential for vessel detection.
This section compares the efficacy of the representation from different self-supervised tasks onto a downstream supervised vessel segmentation task.
We compare our model with four self-supervision baselines, namely,  context-encoder~\cite{pathakCVPR16context}, 
image-denoising~\cite{vincent2008extracting}, image-colorization~\cite{richardECCV2016} 
and Models Genesis~\cite{zhou2019models}.
First, we train multiple networks using different pretext tasks on the DRIVE dataset.
These networks are then \textit{finetuned} on a supervised vessel segmentation task on the STARE dataset.
We consider a \textit{limited-data} and a \textit{high-data} scenario, where finetuning is done with only 4 and 16 images respectively.
Figure \ref{fig:selfsupervised}(a,b) show the training dynamics in both cases.
Our method takes fewer iterations to converge compared to the other methods and achieves the best validation dice score.
\subsection{Segmentation around Bifurcation Points (BPs)}

\begin{figure}[!t]
    \centering
    \begin{minipage}{0.91\textwidth}
    \includegraphics[width=0.19\textwidth]{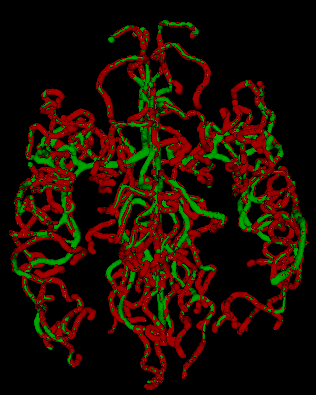}
    \includegraphics[width=0.19\textwidth]{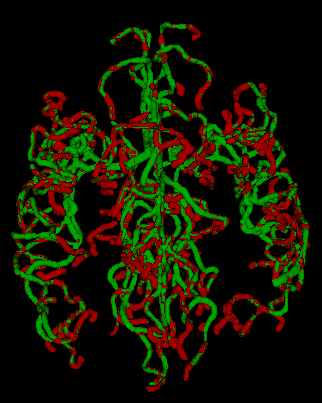}
    \includegraphics[width=0.19\textwidth]{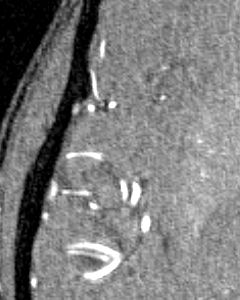}
    \includegraphics[width=0.19\textwidth]{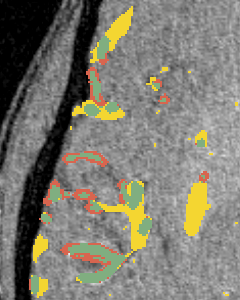}
    \includegraphics[width=0.19\textwidth]{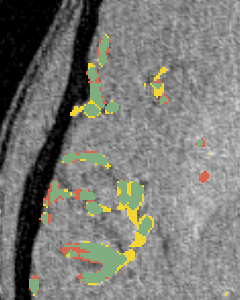}

    \includegraphics[width=0.19\textwidth]{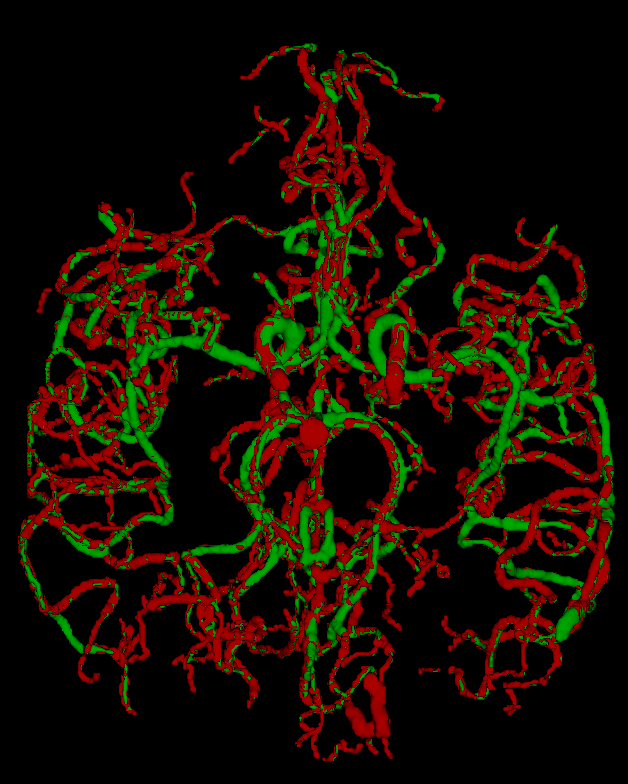}
    \includegraphics[width=0.19\textwidth]{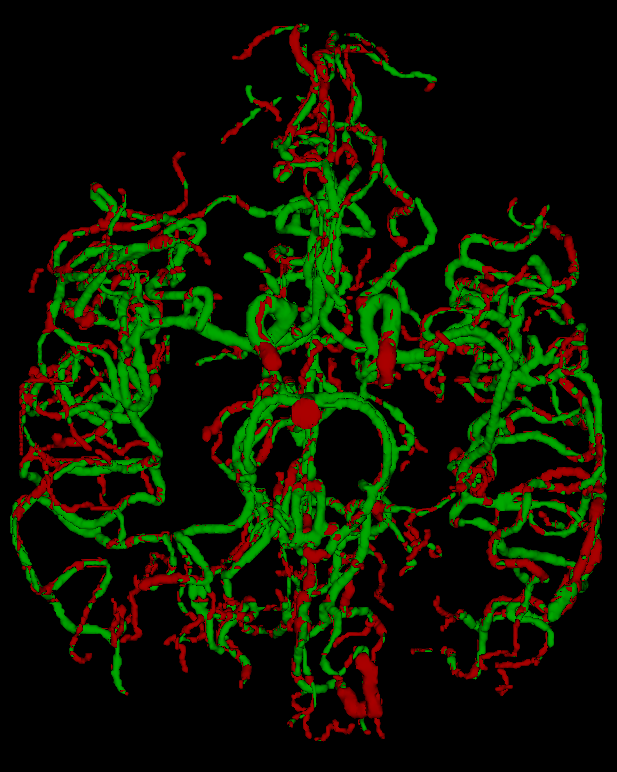}
    \includegraphics[width=0.19\textwidth]{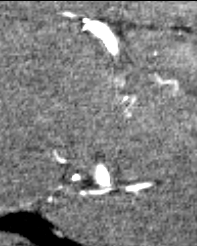}
    \includegraphics[width=0.19\textwidth]{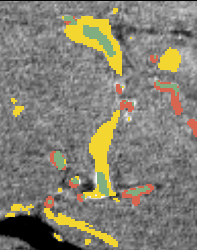}
    \includegraphics[width=0.19\textwidth]{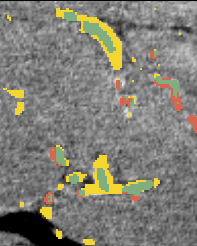}
    \end{minipage}
    
    \resizebox{0.9\textwidth}{!}{
        \begin{minipage}{0.19\textwidth}
        \centering (a)
        \end{minipage}
        \begin{minipage}{0.19\textwidth}
        \centering (b)
        \end{minipage}
        \begin{minipage}{0.19\textwidth}
        \centering (c)
        \end{minipage}
        \begin{minipage}{0.19\textwidth}
        \centering (d)
        \end{minipage}
        \begin{minipage}{0.19\textwidth}
        \centering (e)
        \end{minipage}
    }
    
    \caption{Results on the TubeTK dataset. \textbf{(a)} Result of Frangi segmentation \textbf{(b)} Result of our segmentation (green denotes true positives and red denote false negatives). \textbf{(c)} Axial slice of the image containing vascular structure \textbf{(d)} Result of Frangi segmentation \textbf{(e)} Result of our segmentation. (Green denotes true positives, yellow denotes false positives and red denotes false negatives.) Our method has fewer yellow and red regions.}
    \label{fig:tubetk}
\end{figure}

In this experiment, we demonstrate the importance of our predicted bifurcation flow fields ($\vb_1, \vb_2$) in vessel-segmentation performance around BPs. 
To quantitatively measure the segmentation, we manually annotated bounding boxes (BBs) at multiple bifurcation regions in the DRIVE dataset.
We performed an ablation study, where we didn't consider bifurcation loss (BL) in our final formulation.
Figure \ref{fig:selfsupervised}(c) reports the accuracy of identifying vessel pixels within the extracted BBs.
Our proposed bifurcation loss significantly improves the segmentation at regions around BPs.

%% file: sections/conclusion_v1_s.tex
\section{Conclusion}
Our proposed self-supervised model demonstrates the ability to perform efficient vessel-segmentation on real 2D digital retinal images and 3D CT and MRA scans. 
It does so by using critical properties of tube-like structures such as connectivity, profile consistency, and bifurcations to introduce inductive bias into deep learning and learn in a self-supervised setting. 
Our adaption of self-supervised task demonstrates robustness and generalizability in features in a downstream segmentation task.
To summarize, our work is a step towards incorporating geometrical constraints of tube-like structures into a deep learning framework and providing a robust self-supervised model for vessel segmentation. 
Further work should explore improving the segmentation performance on thin, low contrast vessels.  
A prospective study may explore employing the vessel-segmentation to understand a disease manifestation and establishing clinical usage. 

\section{Acknowledgment}
This work was partially supported by NIH Award Number 1R01HL141813-01, NSF 1839332 Tripod+X, SAP SE, and Pennsylvania Department of Health. We are grateful for the computational resources provided by Pittsburgh SuperComputing grant number TG-ASC170024.

%% file: sections/appendix_v1.tex
\section*{Supplementary Material}

\begin{figure}[ht!]
    \centering
    \includegraphics[width=0.987\textwidth]{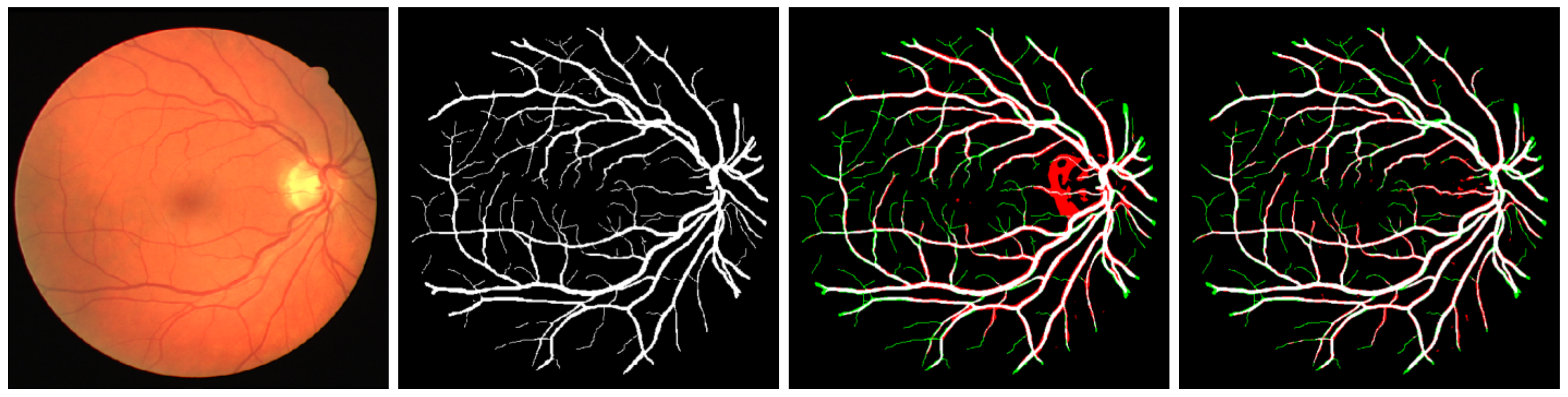}
    \includegraphics[width=0.987\textwidth]{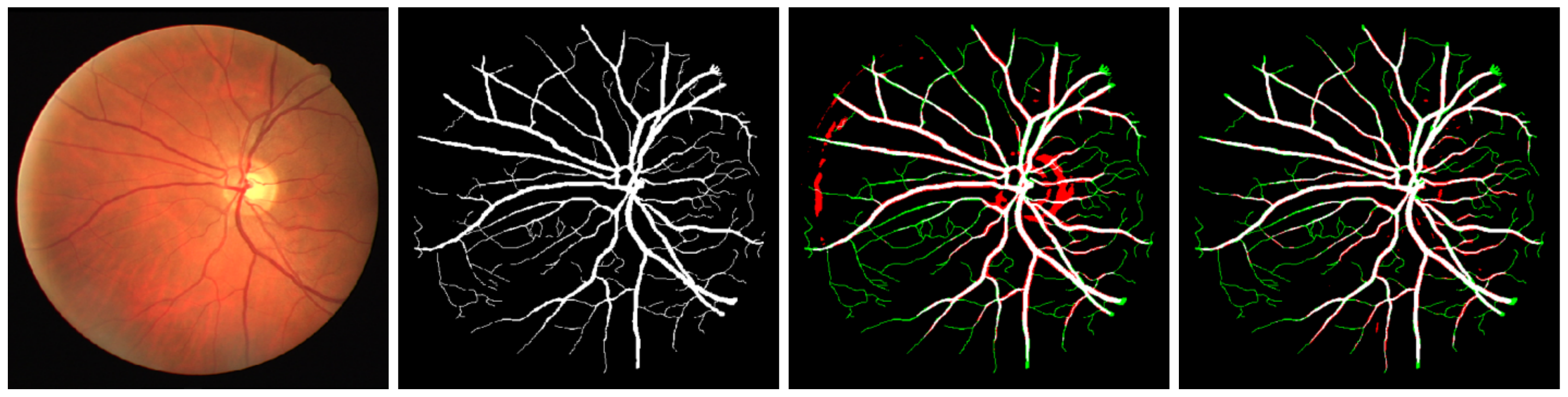}
    \includegraphics[width=0.98\textwidth]{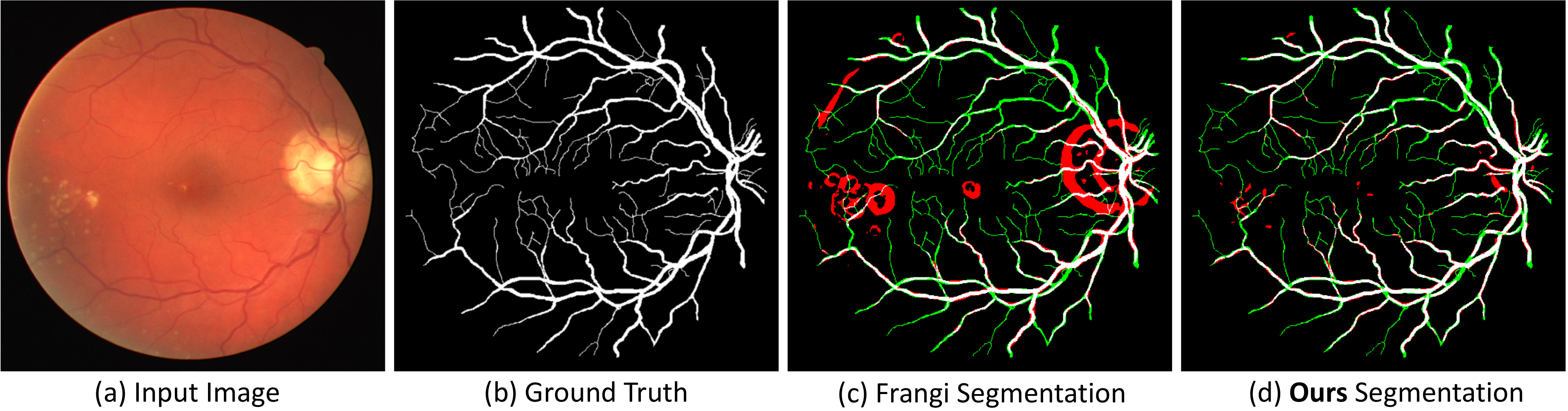}
    \caption{Qualitative results on DRIVE dataset. Different colors are used to highlight true positives (white), false positives (red), and false negatives (green). Frangi segmentation misclassifies ridges as vessels resulting in high false positives (indicated by the presence of big blobs of red regions in figure (c)). Both methods fail to detect very thin vessels, and improving performance for small vessels is an avenue for future work.}
    \label{fig:drive}
\end{figure}
